\documentclass[a4paper,twoside]{article}
\newcommand{\affil}[1]{$^{\rm #1}$}
\usepackage[authoryear]{natbib}
\bibpunct{(}{)}{;}{a}{}{,}
\usepackage{graphicx}
\date{} 
\def\ltsima{$\; \buildrel < \over \sim \;$}
\def\simlt{\lower.5ex\hbox{\ltsima}}
\def\gtsima{$\; \buildrel > \over \sim \;$}
\def\simgt{\lower.5ex\hbox{\gtsima}}
\def\kms{{\rm\,km\,s^{-1}}}

\def\AA{\; \buildrel \circ \over {\rm A}}
\def\AA{$\; \buildrel \circ \over {\rm A}$}

\def\deg{^\circ}

\def\s{\ifmmode \widetilde \else \~\fi}
\def\={\overline}

\def\spose#1{\hbox to 0pt{#1\hss}}

\def\lta{\mathrel{\spose{\lower 3pt\hbox{$\mathchar"218$}}
     \raise 2.0pt\hbox{$\mathchar"13C$}}}
\def\gta{\mathrel{\spose{\lower 3pt\hbox{$\mathchar"218$}}
     \raise 2.0pt\hbox{$\mathchar"13E$}}}
\def\Dt{\spose{\raise 1.5ex\hbox{\hskip3pt$\mathchar"201$}}}    
\def\dt{\spose{\raise 1.0ex\hbox{\hskip2pt$\mathchar"201$}}}    

\def\dotsfill{\leaders\hbox to 1em{\hss.\hss}\hfill}

\title{\large\bf\flushleft Correcting the influence of an asymmetric line spread function in 
2-degree Field spectrograph data}
\author{
\parbox{\textwidth}{\flushleft
\vspace{-0.5cm}
{\it Nicolas F. Martin\affil{A,D}, Rodrigo A. Ibata\affil{A}, Blair C. Conn\affil{B}, Mike J. Irwin\affil{C}, and
Geraint
 F. Lewis\affil{B}}\\
\vspace{0.4cm}
{\small \affil{A}\,Observatoire de Strasbourg, 11 Rue de l'Universit\'e, F-67000 Strasbourg, France}\\
{\small \affil{B}\,Institute of Astronomy, School of Physics, A29, University of Sydney, NSW 2006, Australia}\\
{\small \affil{C}\,Institute of Astronomy, Madingley Road, Cambridge, CB3 0HA, U.K.}\\
{\small \affil{D}\,Email: martin@astro.u-strasbg.fr}}}
\begin{document}
\twocolumn[
\begin{changemargin}{.8cm}{.5cm}
\begin{minipage}{.9\textwidth}
\vspace{-1cm}
\maketitle
%
%
\small{\bf Abstract:} We investigate the role of asymmetries in the line spread function of the 2-degree field
spectrograph and the variations in these asymmetries with the CCD, the plate, the time of observation and the fibre. A
data-reduction pipeline is developed that takes these deformations into account for the calibration and cross-correlation of the
spectra. We show that, using the emission lines of calibration lamp observations, we can fit the line spread function
with the sum of two Gaussian functions representing the theoretical signal and a perturbation of the system. This model
is then used to calibrate the spectra and generate templates by downgrading high resolution spectra. Thus, we can
cross-correlate the observed spectra with templates degraded in the same way. 

Our reduction pipeline is tested on real observations and provides a significant improvement in the accuracy of the
radial velocities obtained. In particular, the systematic errors that were as high as $\sim 20\kms$ when applying the
AAO reduction package 2dfdr are now reduced to $\sim5\kms$.

Even though the 2-degree Field spectrograph is to be decommissioned at the end of 2005, the analysis of archival data and previous
studies could be improved by the reduction procedure we propose here.

\medskip{\bf Keywords: instrumentation: spectrograph -- techniques: spectroscopic -- galaxies: kinematics and dynamics}

\medskip
\medskip
\end{minipage}
\end{changemargin} ]
\small

\section{Introduction} The 2-degree field multi-object spectrograph at the Anglo-Australian Telescope can observe up to
400 objects within a 2 degree field on the sky through 2 different spectrograph settings at the same time
\citep{lewis02}\footnote{see also {\tt http://www.aao.gov.au/2df/}}.  Each spectrograph accepts light from 200 fibres,
which are positioned by a robot on a plate in the focal plane of the telescope, to produce spectra of low to medium
resolution depending on the chosen grating. Two full sets of fibres on separate plates ensures that re-configuring can
be done in parallel with observing, hence minimizing time lost to the positioning of fibres during the night. 

With its high resolution gratings, 2dF should be an ideal instrument for studies of the kinematics of resolved stars,
however, it has been shown that radial velocities obtained using the 2dF spectrograph are plagued by systematic errors
of the order of $10-20\,\kms$ \citep{cannon02,stanford02}. While known, these systematic errors are not easily corrected 
for they are not constant over all the observations. Observations performed on different CCDs or plates, during
different nights or at different times during the same night can show variations in radial velocity that are worse than
the $5\,\kms$ internal precision the 2dF should achieve.

The most probable culprit for these variations is a change in the point spread function of the system across the CCDs
due the optics of the cameras, tilts in the slit blocks or changes in the focus of the spectrographs with temperature or
orientation of the telescope, given that they are mounted on the top of the AAT close to the dome aperture
\citep{cannon02}. These deformations produce an asymmetric Line Spread Function (LSF) on the extracted spectra which can
be revealed by artificially asymmetric emission/absorption lines.

Our group recently undertook a radial velocity survey of the region around the Canis Major (CMa) dwarf galaxy and other
nearby overdensities in Red Giant Branch stars (RGBs) and Red Clump stars \citep{martin04,martin05}. Given its high
number of targets on a wide field of view, the 2dF spectrograph is currently the best instrument for such a study.
However, one of the key features of accretion streams is their very low dispersion in radial velocities. Thus, it is
crucial that we achieve the lowest possible systematic errors on the velocities of our sample stars to be as close as
possible to the internal dispersion of a population. Moreover, for the case of the Canis Major study, the velocity of
the constituent stars is close to those of the thin disc \citep{martin04}. In the regions where the CMa population is
significantly contaminated by disc stars, a systematic error of $10-20\,\kms$ could diffuse the signal of the
extra-galactic population and could prevent detection.

For these reasons, we constructed a reduction pipeline for 2dF observations that takes into account the asymmetry of the
LSF of the extracted spectra and corrects it. Since the shape of the LSF plays a role in the calibration of the observed
spectra and their cross-correlation with templates to obtain radial velocities, our pipeline works as follows:

\begin{enumerate}
\item determination of the deformations of the LSF across the CCD using the emission lines of the observed calibration
lamps;
\item use of the determined model of the LSF to calibrate the arc spectra;
\item use of the determined model of the LSF to generate template spectra (from high resolution spectra) deformed in the
same way as the observed spectra;
\item cross-correlation of the observed spectra with the obtained template spectra.
\end{enumerate}
 
In section 2, we describe the different steps of the pipeline, while in section 3 the derived radial velocities are
compared to reference observations and we analyze the accuracy of the new velocities. Section 4 summarizes the pipeline
and the results.

In the following, the 2dF observations are first corrected for the flat field and extracted using the 2dF Data Reduction
package (2dfdr) provided by the AAO \citep{taylor96}\footnote{See also {\tt
http://www.aao.gov.au/2df/software.html\#2dfdr}}. Sky subtraction is also performed using the 2dfdr package.

\section{ Observations and Reduction}

The observations on which we apply our reduction pipeline in this section are part of our Canis Major radial velocity
survey. We targeted the Red Giant Branch stars of a field centred on $l=240\deg$, $b=-8.8\deg$.
We employed two different   spectrograph   settings,  with   the   1200V  grating  
on spectrograph  1 (covering  4600--5600\AA\ at  1\AA/pixel, these are the CCD1 data) and with the 1200R grating on
spectrograph 2 (covering 8000--9000\AA, also at  1\AA/pixel, these are the CCD2 data).

\citet{stanford02} showed that observations made with CCD2 are better than those obtained through CCD1. We will
therefore concentrate of correcting the deformations of the LSF on CCD1. This is also justified by our analysis of CCD2
observations (see below, subsection 2.5).

\subsection{The LSF model}

For a given optical system, the output signal $s_o(\lambda)$ that is observed at wavelength
$\lambda$ can be expressed as: 

\begin{equation}
s_o(\lambda)=(s_i\ast G)(\lambda)
\end{equation}

\noindent where $s_i$ is the input of the system and $G$ is the line spread function of the system. When observed on a
CCD, each pixel $X$, of width $S$\AA\ receives part of this output signal, and has a value $F(X)$ such that:

\begin{equation}
F(X)=\int_{X-0.5\cdot S}^{X+0.5\cdot S}s_o(\lambda)\,\textrm{d}\lambda \, .
\end{equation}

When observing an Emission Line (EL) centred on $\mu_1$, the input signal is:

\begin{equation}
s_i(\lambda)=\delta(\lambda-\mu_1)
\end{equation}

\noindent which produces the observed signal:

\begin{equation}
s_o(\lambda)=\int_{-\infty}^{\infty}G(\tau)\delta(\lambda-\mu_1-\tau)\,d\tau=G(\lambda-\mu_1) \, .
\end{equation}

\noindent This means that the signal, $F(X)$, produced by this EL on pixel $X$ of the 2dF with grating 1200V --- for 
which $S=1$\AA\ --- is:
\begin{equation}
\label{EL} F(X)=\int_{X-0.5}^{X+0.5}G(\lambda-\mu_1)\,\textrm{d}\lambda \, .
\end{equation}

If the system through which the observations are performed was perfect, the LSF of the system ($G$) would be a symmetric
Gaussian function with a dispersion that depends on the quality of the system. However, this is not the case for the 2dF
spectrograph. As can be seen on Figure~1 for an EL produced by the observation of a Copper-Argon calibration lamp on
CCD1, the LSF is deformed and generates an EL with an asymmetric wing at higher wavelengths. This deformation is
significant and can account for as much as 30\% of the total signal of the EL. 

\begin{figure}[t]
\begin{center}
\includegraphics[width=0.7\hsize, angle=270]{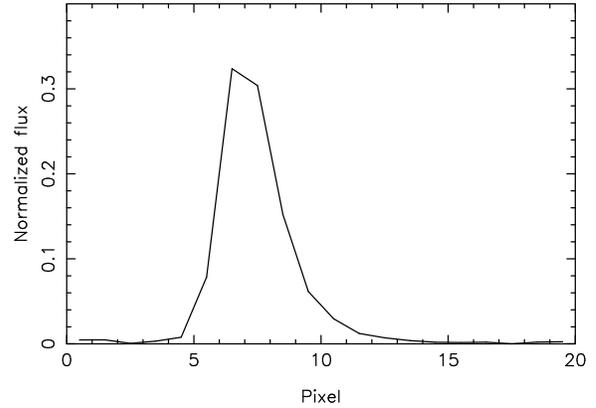}
\caption{5105.541\AA\ Cu emission line extracted from fibre 1 of a CCD1 calibration lamp observation. The emission line
shows an asymmetry towards higher wavelengths that accounts for $\sim30$~\% of the total signal.}
\end{center}
\end{figure}

To model the deformed LSF, we represent it as the sum of two Gaussian functions. The first one, $G_1$, is the signal that
would be observed through a perfect system and is only determined by the dispersion of the would be perfect system:
$\sigma_1$. The second Gaussian function, $G_2$, is considered as a perturbation of the system that produces the
deformation. It is shifted by $\Delta\mu$ compared to $G_1$, has its own dispersion $\sigma_2$ and contains a fraction
$A$ of the total signal. Hence this model of the LSF can be expressed by:

{\setlength\arraycolsep{2pt}
\begin{eqnarray}
\label{LSF} G(\lambda) & = & (1-A)\,G_1(\lambda)+A\,G_2(\lambda)\nonumber\\
& = & \frac{1-A}{\sqrt{2\pi}\,\sigma_1}e^{-\frac{1}{2}\big(\frac{\lambda}{\sigma_1}\big)^2}+
\frac{A}{\sqrt{2\pi}\,\sigma_2}e^{-\frac{1}{2}\big(\frac{\lambda-\Delta\mu}{\sigma_2}\big)^2}.
\end{eqnarray}

With this model and using equation~\ref{EL}, the value $F(X)$ of pixel $X$ of an emission line becomes:

{\setlength\arraycolsep{2pt}
\begin{eqnarray}
\label{EL2} F(X) & = &\frac{1-A}{\sqrt{2\pi}\,\sigma_1}\int_{X-0.5}^{X+0.5}
e^{-\frac{1}{2}\big(\frac{\lambda-\mu_1}{\sigma_1}\big)^2}\,\textrm{d}\lambda\\
& & +\frac{A}{\sqrt{2\pi}\,\sigma_2}\int_{X-0.5}^{X+0.5}
e^{-\frac{1}{2}\big(\frac{\lambda-(\mu_1-\Delta\mu)}{\sigma_2}\big)^2}\,\textrm{d}\lambda.\nonumber
\end{eqnarray}
}

\subsection{Fitting the model}

As expressed in equation~\ref{EL2}, the shape of an EL at a given position on the CCD is
only determined by 5 free parameters:

\begin{itemize}
\item $\mu_1$: the theoretical centre of the emission line;
\item $\sigma_1$: the standard deviation of the theoretical, symmetric LSF;
\item $A$: the fraction of the total signal that is in the perturbation;
\item $\sigma_2$: the standard deviation of the perturbation;
\item $\Delta\mu$: the shift between the two Gaussian functions $G_1$ and $G_2$. Since the asymmetry of the emission
lines is to higher wavelengths, $\Delta\mu>0$.
\end{itemize}

$\mu_1$ is the only free parameter that depends on the EL while the 4 others ($A$, $\sigma_1$, $\sigma_2$ and
$\Delta\mu$) define the shape of the LSF for the considered fibre, time and plate. By fitting this model $F$ of the EL
on the EL observed on the CCD, $\widetilde{F}$\footnote{In the following, the $^\sim$ functions (e.g. $\widetilde{F}$)
represent the observed data while the functions without the $^\sim$ represent the model to be fitted to the data (e.g.
$F$).} (such as in Figure~1), we can deduce the parameters that define the LSF of the system for the fibre, wavelength,
time and plate of the observation.

To completely characterize the deformations of the LSF, it would be necessary to define these parameters in both the X
(wavelength) and Y (fibre number) directions of the CCD. However, since for our observations the region we use to
cross-correlate the spectra and the templates is near the centre of the wavelength range of the CCD and since there
are few ELs in this region, we only model the deformations of the LSF in the Y direction of the CCD. Yet, observations
performed with different settings may require a complete two dimension model of the LSF.

Among the $\sim 10$ ELs that are observed with the 1200V grating with a Copper-Argon calibration lamp, the best choice
to fit the model $F$ is the Cu line at 5105.541\AA\ since it is the strongest line and it is at the centre of the region
we will use to cross-correlate the spectra and the templates (the 4800\AA\ to 5250\AA\ region)\footnote{We choose this
region for the cross-correlation with the templates because it is free of strong sky features. It is also located at the
centre of the CCD along the wavelength axis, which should diminish potential asymmetries of the LSF generated along this
axis.}. We could determine the 5 free parameters of the model on the $\widetilde{F}$ function of each fibre, however,
the low number of pixels over which $\widetilde{F}(X)\simgt0.0$ (less than 20, see Figure~1) would produce high
uncertainties on the parameters. Therefore, we choose to fit the same LSF model on up to 10 consecutive fibres
(depending on the presence of dead fibres) at the same time. Indeed, the spectrograph is constructed with fibres coming
in blocks of 10 which means the fibres in each of these blocks should have analogous LSF\footnote{This was tested by
taking the fibres in smaller groups. The results are similar, but with higher uncertainties on the values of the
parameters.} but changes may appear between each block.

\begin{figure*}
\includegraphics[width=4.2in,angle=270]{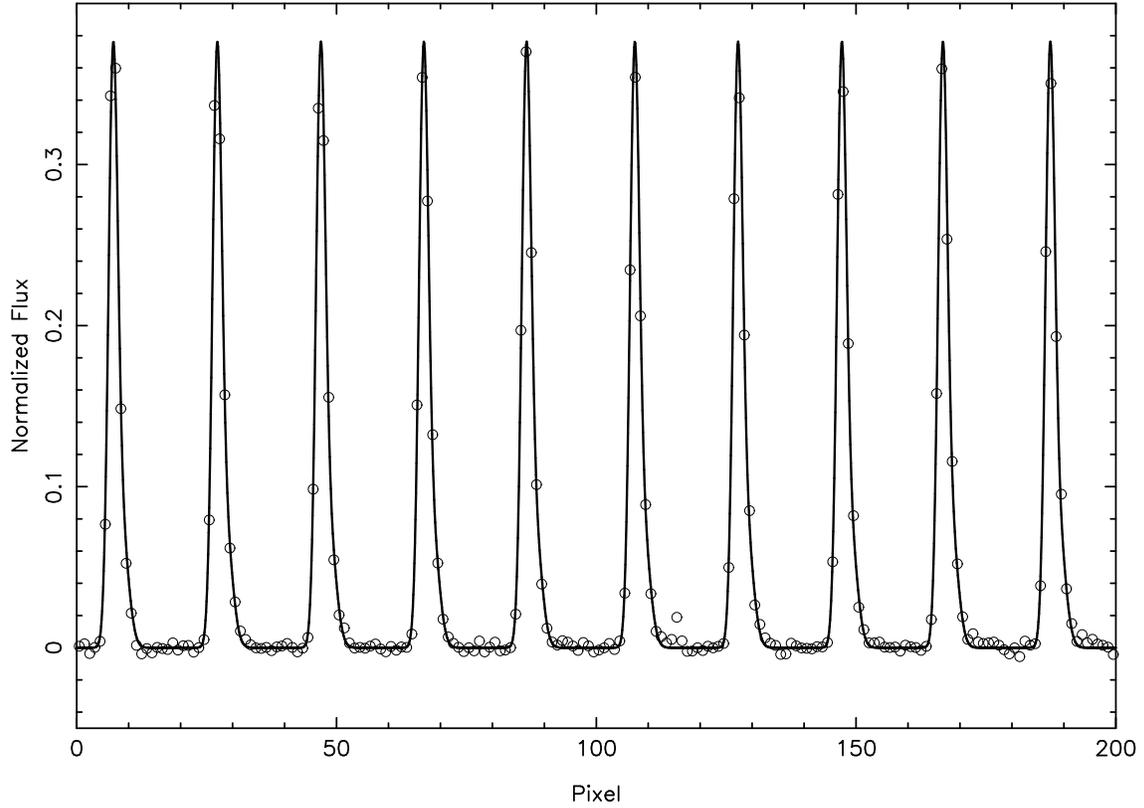}
\caption{Concatenation of emission lines from a group of fibres (hollow circles, corresponding to the $\widetilde{F'}$
function). Since there are no dead fibres in this group, 10 emission lines were extracted and were used for the fit
following the procedure described in the text. The best fit (the $F'$ function) is shown as a thick line and faithfully
represents the data. In particular, it reproduces the asymmetric, high wavelength wing of each emission line.}
\end{figure*}

By extracting the EL of all the fibres and concatenating them, we generate a new function $\widetilde{F'}$ that is
modeled by $F'$:

\begin{equation}
F'(X)=\sum_{i=1}^{N}\int_{X-0.5}^{X+0.5}G(\lambda-\mu_i)\,\textrm{d}\lambda
\end{equation}

\noindent In this new model, $G$ is still the LSF model as expressed in equation~\ref{LSF}, $\mu_i$ is the centre of the
$i^{\mathrm{th}}$ EL and $N$ is the number of fibres in the group (i.e. 10 minus the number of dead fibres). This
concatenated model $F'$ now contains $4+N$ free parameters: the $N$ centres of the ELs $(\mu_i)_{1\le i\le N}$ and the 4
parameters of the LSF ($A$, $\sigma_1$, $\sigma_2$ and $\Delta\mu$). This time however, there are $20\times N$ points to
adjust these $4+N$ free parameters which ensures lower uncertainties on the estimate of the parameters.

Each group of fibres (1-10, 11-20, \dots) is processed as follows:

(i) for each fibre:
\begin{itemize}
\item check if fibre is not dead;
\item extract the Cu line at 5105.541\AA\ (around pixel 500) to generate $\widetilde{F}$;
\end{itemize}

(ii) concatenate all the fibres of the group to generate $\widetilde{F'}$;

(iii) fit model $F'$ ($4+N$ parameters) to data $\widetilde{F'}$.

When extracting the EL, we take care to extract approximately the same region around the line for all the fibres, even
if their centre varies with the position across the CCD. We detect the pixel $X_{\mathrm{C}}$ with the highest value
of the EL and extract 6 pixels before $X_{\mathrm{C}}$ and 13 pixels after $X_{\mathrm{C}}$ to account for the
asymmetry to higher wavelength in the shape of the EL. We then correct from a possible continuum by subtracting a linear
fit between the average value of the first 3 pixels of $\widetilde{F}$ and the average value of the last 3 pixels.
Finally, to correct differences in the response of the fibres, we normalize each EL so that:

\begin{equation}
\int\widetilde{F}(X)\,\textrm{d}X=1.0
\end{equation}

\noindent An example of the final data $\widetilde{F'}$ we obtain after the concatenation of the fibres of a group is
shown as hollow circles on Figure~2.

To fit the model $F'$ to the data $\widetilde{F'}$, we use the {\tt mrqmin} routine described in \citet{press92} and
determine the $4+N$ parameters. Since determining all of the parameters at the same time proves difficult, we determine
independently the $N$ EL-related parameters $(\mu_i)_{1\le i\le N}$ and the 4 LSF-related parameters $A$, $\sigma_1$,
$\sigma_2$ and $\Delta\mu$. The fitting procedure is as follows:

(i) the $N$ $\mu_i$ values are determined for a loose grid of fixed values of $A$, $\sigma_1$, $\sigma_2$ and
$\Delta\mu$. The $(\mu_i)_{1\le i\le N}$ producing the lowest $\chi^2$ (in the sense of {\tt mrqmin}) are used as input
values;

(ii) the $(\mu_i)_{1\le i\le N}$ are fixed and the 4 LSF-related parameters are determined;

(iii) these LSF-related parameters are fixed to their best values and used as input values to re-determine the
$(\mu_i)_{1\le i\le N}$;

(iv) these are once again fixed to re-determine the best values of $A$, $\sigma_1$, $\sigma_2$ and $\Delta\mu$.

Steps (iii) and (iv) are used as a sanity check to verify the independence of the two sets of parameters. Since $\chi^2$
is reduced by less than 1\% between step (ii) and step (iv), our hypothesis can be taken as valid and the 2 sets of
parameters determined independently. Figure~2 shows the fit resulting from this procedure as a thick line. It is
visibly good with a well-determined position for the ELs (the different $\mu_i$) and it reproduces the asymmetric shape
of the ELs with a large wing at higher pixel number.

\begin{figure}[t]
\begin{center}
\includegraphics[width=0.7\hsize, angle=270]{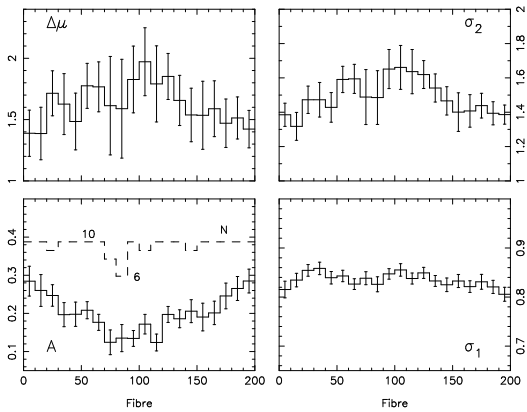}
\caption{Evolution of the 4 LSF-related parameters $A$ (bottom left panel), $\sigma_1$ (bottom right), $\sigma_2$ (top
right) and $\Delta\mu$ (top left) across CCD1. The three parameters representing the perturbation of the LSF ($A$,
$\sigma_2$ and $\Delta\mu$) all show substantial variations across the CCD. The number of fibres, $N$, that were used
for the fit is plotted as a thin dashed histogram in the bottom left panel.}
\end{center}
\end{figure}

\begin{figure*}[t]
\includegraphics[width=0.385\hsize,angle=270]{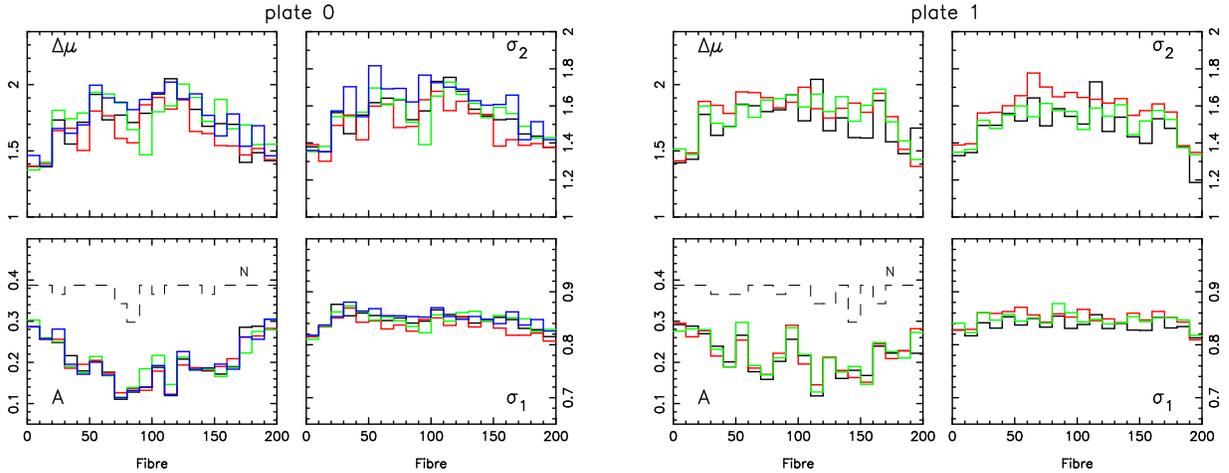}
\caption{Same as Figure~3 but with the evolution of the LSF parameters for all the observations that were made during a
single night. The parameters for observations on plate 0 are shown on the left panels and those performed on plate 1 are
shown on the right panels. Altough the parameters evolve in a similar way for all the fits, there are variations with
plate and/or observation that prevent from using a generic set of parameters for the whole night.}
\end{figure*}

For a given calibration lamp observation, this procedure is repeated for the 20 groups of up to 10 valid fibres. An
example of the evolution of the 4 LSF-related parameters is shown on Figure~3. The first striking feature is that the 4
parameters evolve across the CCD, which means that the asymmetry of the LSF is not constant for the different fibres of
the same observation run. The most important parameter, $A$, which represents the weight of the perturbation goes up to
$\sim30\%$ at the edges of the CCD but is as low as $\sim15\%$ at its centre. If this behaviour is expected
\citep{cannon02}, the deformations at the edges of the CCD are so important they can easily explain strong systematic
errors in the radial velocities if they are not accounted for. The parameters all have a roughly symmetric evolution
with values at the two edges of the CCD being in the same range and different from those in the centre. 

One of the assumptions in our LSF model is that it can be divided into a ``theoretical signal'' $G_1$ (determined by
$\sigma_1$) and a ``perturbation'' $G_2$ (determined by $A$, $\Delta\mu$ and $\sigma_2$) that creates the asymmetry.
This seems to be confirmed by the almost constant value of $\sigma_1$ that only evolves between 0.81 and 0.85 over the
CCD, while at the same time, the ``perturbation dispersion'', $\sigma_2$, is in a range that is one order of magnitude
higher (between 1.3 and 1.7). Moreover, the value of $\sigma_1$ is similar to what is found on the symmetric ELs
observed on CCD2 (see below, subsection 2.5).

Though the parameters evolve in the expected way, a more detailed look at their behaviour shows many gaps and spikes
where one would expect a smoother evolution over the CCD. This is partly due to changes in the number of fibres used to
fit the model ($N$, dashed histogram in the bottom left panel of Figure~3): for instance, there are four dead fibres
between fibre 81 and fibre 90 so, in the corresponding group, there remains only 6 ELs for the fit. For this group, the
parameters $\Delta\mu$ and $\sigma_2$ show a gap and have wider error bars. These gaps and spikes are also explained by
the shallow $\chi^2$ space which provides many slightly different LSF models with small differences in $\chi^2$.

Another interesting point to look at is the evolution of the asymmetry on the LSF during a night of observation and the
influence of the plate on which the observations are done. Figure~4 shows the evolution of the 4 parameters of the LSF
for the 7 observation runs we made on 9th April 2004. For the same plate, the results are similar, with parameters that
are really identical (e.g. $\sigma_1$ and $A$) but the parameters of the ``perturbation'' tend to vary, even though they
show a similar behaviour. Likewise, even if the deformation of the LSF evolves in a similar way on the two plates, there
are some intrinsic differences between the two: for instance, the amplitude of the deformation tends to be higher at the
center of the CCD on plate 1, the standard deviation of the ``perturbation'', $\sigma_2$, tends to be offset to lower
values.

Therefore, even if it is reassuring to observe a similar behaviour in the evolution of the deformation of the LSF during
the night, it seems preferable to determine a LSF model on the calibration lamp of each observation run. Using
a global LSF model for all the observations done on a single plate during a single night nevertheless yields acceptable
final results, even though the uncertainty of the derived radial velocities tend to be 20-30 percent higher than when
using one LSF model for each calibration lamp.


\subsection{Calibration}

One part of the reduction where the asymmetry of the LSF has an important effect is the calibration of the spectra.
Indeed, in a routine like {\tt identify} in IRAF the centre, $\mu_0$, of an emission line $\widetilde{F}$ defined over
the range $[X_{\mathrm{min}},X_{\mathrm{max}}]$ is the wavelength at which the EL can be divided in two equally luminous
parts. That is:

\begin{equation}
\label{mu0}
\int_{X_{\mathrm{min}}}^{\mu_0}\widetilde{F}(X)\,\textrm{d}X
=\int_{\mu_0}^{X_{\mathrm{max}}}\widetilde{F}(X)\,\textrm{d}X.
\end{equation}

\begin{figure}[t]
\begin{center}
\includegraphics[width=0.7\hsize, angle=270]{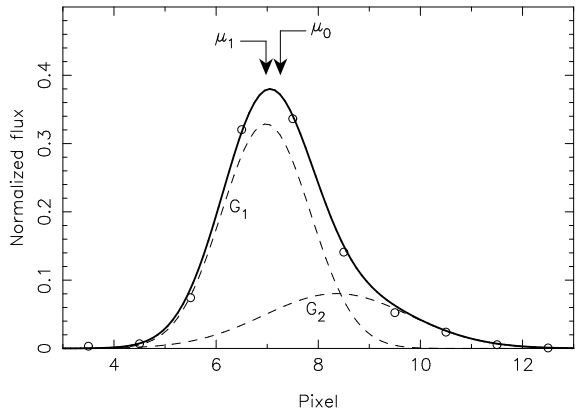}
\caption{Difference in the central position of an EL for two calibration methods. $\mu_0$ is defined as in
equation~\ref{mu0} and corresponds to the value given by the IRAF routine {\tt identify} while $\mu_1$ is the centre of
the ``theoretical signal'', $G_1$, of the LSF, as defined in equation~\ref{LSF}. The observed values, $F$, of the EL are
shown as hollow circles, while the LSF model, $\widetilde{F}$, determined following the procedure of subsection 2.2 is
represented as a thick line. Its two components (the ``theoretical signal'' $G_1$ and the ``perturbation'' $G_2$)
are plotted as thin dashed lines. In this example, there is a shift of 0.28\,pixels between the centres of the LSF,
$\mu_0$ and $\mu_1$, determined by the two different methods.}
\end{center}
\end{figure}

\noindent Hence, the presence of the asymmetry in the EL tends to switch the ``IRAF centre'' $\mu_0$ to higher values
than the ``theoretical centre'' $\mu_1$ which should be considered as a more valid position of the EL (see Figure~5).
With the evolution of this offset directly related to the LSF parameter $A$, the difference $\mu_0-\mu_1$ changes with
the fibre. And since we cross-correlate the observed spectra with templates that we generate (see next subsection) and
do not have this calibration issue, the deduced radial velocity can have a systematic error of up to $\sim 10\,\kms$.

To correct from this effect, we have to calibrate each arc spectrum using the LSF shape of the corresponding fibre, time
and plate to determine the position of the ``theoretical'' EL ($\mu_1$) instead of the centre of the asymmetric,
observed, EL ($\mu_0$). For a given fibre of a calibration lamp, we proceed as follows:

(i) for each EL of the corresponding fibre on the calibration lamp, determination of its ``theoretical centre'' $\mu_1$
by fitting the EL model of that fibre (equation~\ref{EL2}) with  $\mu_1$ as the only free parameter (an example is given
Figure~5);

(ii) fit all the ``theoretical centres'' with a Legendre polynomial ($\sim10$ reliable ELs are used for this fit) to
determine the wavelength of each pixel of the observed spectrum;

(iii) the spectrum is re-binned so that each pixel is 1\AA\ wide, starting at 4600\AA\ and ending at 5600\AA.

\begin{figure}[t]
\begin{center}
\includegraphics[width=0.7\hsize, angle=270]{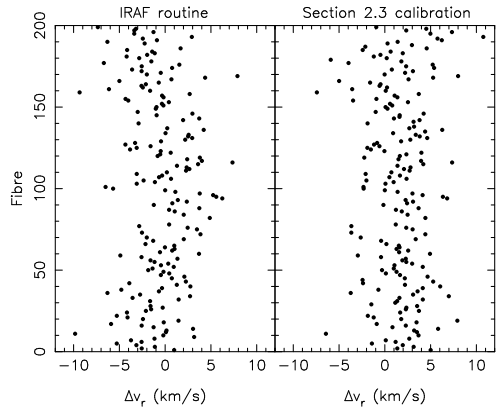}
\caption{Difference in radial velocity between the different fibres and fibre 100 for a twilight observation. On the
left panel, the spectra were calibrated using the IRAF routine {\tt identify} while on the right panel, the calibration
was performed using the shape of the LSF, as explained in subsection 2.3. The IRAF calibration generates an arc shaped
systematic error that depends on the fibre number while our procedure does not introduce such errors.}
\end{center}
\end{figure}

Using the ``theoretical centres'' to calibrate the spectrum ensures a calibration that is less altered by the shape of
the 2dF LSF. Indeed, though we do not take into account possible evolutions of the LSF asymmetry along the wavelength
direction of the CCD --- we fit the same LSF model to all ELs --- this technique gives a much better calibration than
when using the simpler IRAF routine. To test that assumption one can measure the difference in radial velocity for the
different fibres of a twilight observation. Since in such an observation, the spectrum on each fibre is that of the Sun,
cross-correlating each calibrated spectrum with a reference spectrum (say, the one of fibre 100) directly reveals
calibration-related systematic errors. This test was done for twilight observations of 9th April 2004 and the difference
in radial velocity between the different fibres is shown on Figure~6 for a calibration using the IRAF {\tt identify}
routine (left) and the procedure described above (right).

It is clear that directly using the IRAF routine is not satisfactory with a velocity difference between fibre 100 and
the other fibres that is not constant and which follows an arc with increasing discrepancies at the edge of the CCD. On
the other hand, this effect is corrected when using our calibration procedure and all the fibres produce a similar
radial velocity, with a small dispersion of $2.8\kms$.

\subsection{Generating templates}

Measuring the radial velocity of a star requires one to cross-correlate the stellar spectrum with that of a template
star of known radial velocity. Generally, a template spectrum is obtained by observing, along with the target stars, a
few well-known stars. However, due to time constraints, one is forced to observe a small number of templates during a
specific night, with a specific fibre of the 2dF, whereas as we saw earlier, the asymmetry of the LSF can evolve with
time and certainly changes with the fibre (see Figure~4). Consequently, the cross-correlation would be produced by two
spectra deformed in different ways, which can produce systematic errors on the measured shift between the two spectra of
order of 0.2~pixels (corresponding to as much as $15\kms$ on CCD1). Hence, it is necessary to take the LSF shape 
of the templates into account when doing the cross-correlation.

Since it is cumbersome to deconvolve the spectrum of the templates and then reconvolve it with the LSF of the fibre one
is interested in, we choose to generate artificial templates by convolving High Resolution (HR) spectra by the deformed
LSF. This way, the target and template spectra are similarly deformed and their cross-correlation no longer has
systematic effects due to differences in the deformation of the LSF of the two spectra. For each group of 10 fibres, we
proceed as follows:

(i) convolution of the HR template with the corresponding LSF (as defined in equation~\ref{LSF} and determined as
explained in subsection 2.2);

(ii) downgrading the HR, convolved template to a low resolution, 2dF-like template, (1\AA/pix for the 1200V grating);

(iii) cross-correlation of the observed spectra with the artificial template using the {\tt fxcor} routine in IRAF.

We choose the HR spectra from the UVES Paranal Observatory Project\footnote{{\tt 
http://www.sc.eso.org/santiago/uvespop/index.html}}\citep{bagnulo03} which has the advantage of providing the spectrum
of stars of numerous spectral types (from type O stars to type M stars) and hence provides the opportunity to choose a
template that resembles the observed stars. Moreover, the UVES observations made through the 580L grating
(4760--5770\AA) correspond to the range we are interested in (4800--5250\AA).

%

\subsection{CCD2}

\begin{figure}[t]
\begin{center}
\includegraphics[width=0.7\hsize, angle=270]{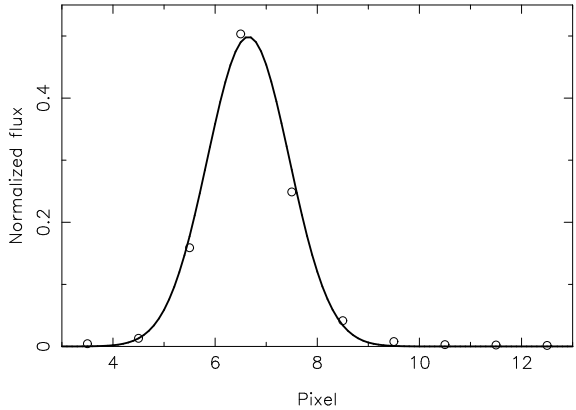}
\caption{8408.2096\AA\ CuAr emission line extracted from fibre 1 of a CCD2 calibration lamp observation (hollow
circles). The emission line is symmetric and can be fitted by a single Gaussian function of internal dispersion
$\sigma=0.74$
(thick line).}
\end{center}
\end{figure}

As is visible on Figure~7, the calibration lamps observed on CCD2 show emission lines that do not have asymmetries like
CCD1. In fact, the emission lines are well fitted by a single Gaussian model (the thick line of Figure~7) with
uncertainties on the position and dispersion of the Gaussian lower than 0.01\,pixel. It is therefore not necessary to
apply the previous reduction techniques to obtain precise radial velocities from observations on CCD2. However, the
symmetry of the emission lines does not mean that there are no issues with this CCD. Indeed, analyzing the same EL for
all the fibres (the 8408.2096\AA\ CuAr line roughly located at the centre of the CCD) shows there are substantial
changes in the internal dispersion, $\sigma$, of the Gaussian fit. Once again, $\sigma$ has an arc-like evolution across
the CCD (see Figure~8). Moreover, the evolution differs with the plate on which the observations were performed. Even
though an increase in the dispersion is expected at the edges of the CCD, the magnitude of the variation (from
$\sigma=0.6$ at the centre up to $\sigma=0.75$ at the edge of plate 0 observations or $\sigma=0.85$ at the edge of plate
1 observations) is troublesome. In addition, there is at least one group of fibres (the 90-100 group on plate 0) that
has a higher internal dispersion than its neighbouring groups.

Altough the symmetry of the ELs means the 2dfdr reduction package should suffice to reduce the data precisely, the
variation of the internal dispersion of the LSF on CCD2 is to be kept in mind if some systematic effects remain at the
end of the reduction.

\begin{figure}[t]
\begin{center}
\includegraphics[width=0.7\hsize, angle=270]{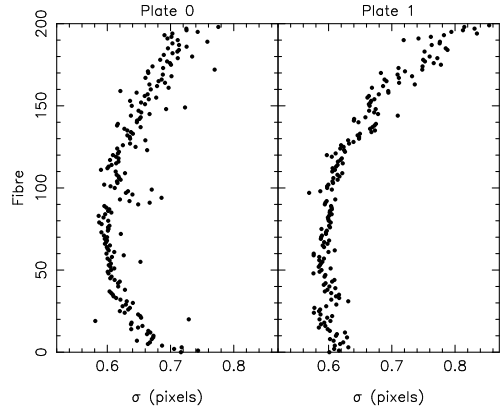}
\caption{Evolution of the internal dispersion, $\sigma$, of the LSF across CCD2 when fitted by a single Gaussian
function as in Figure~7. Although the LSF remains symmetric across CCD2 (contrary to CCD1), its width varies
substantially.}
\end{center}
\end{figure}

\section{Results}

\begin{figure*}[t]
\includegraphics[width=0.385\hsize,angle=270]{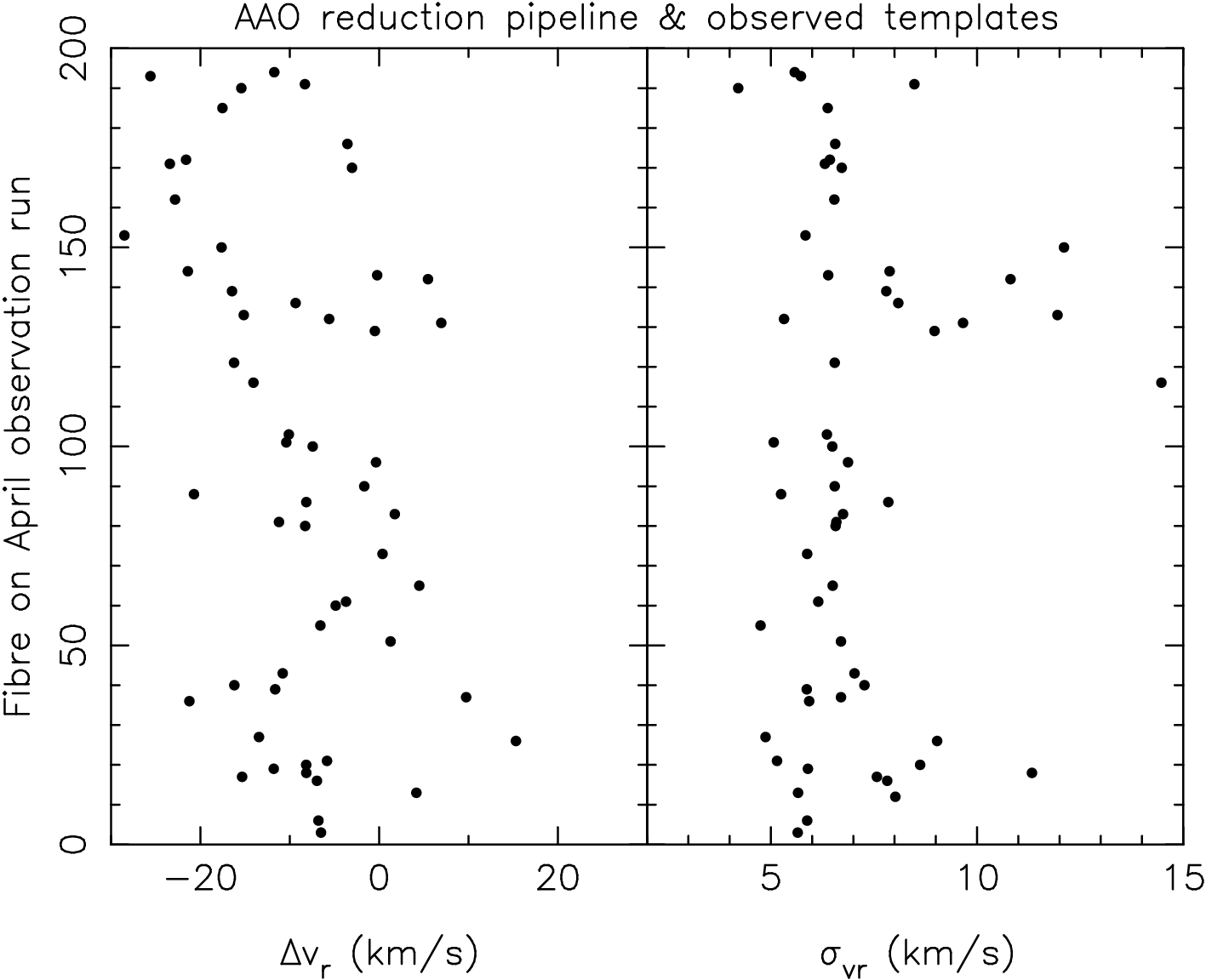}
\includegraphics[width=0.385\hsize,angle=270]{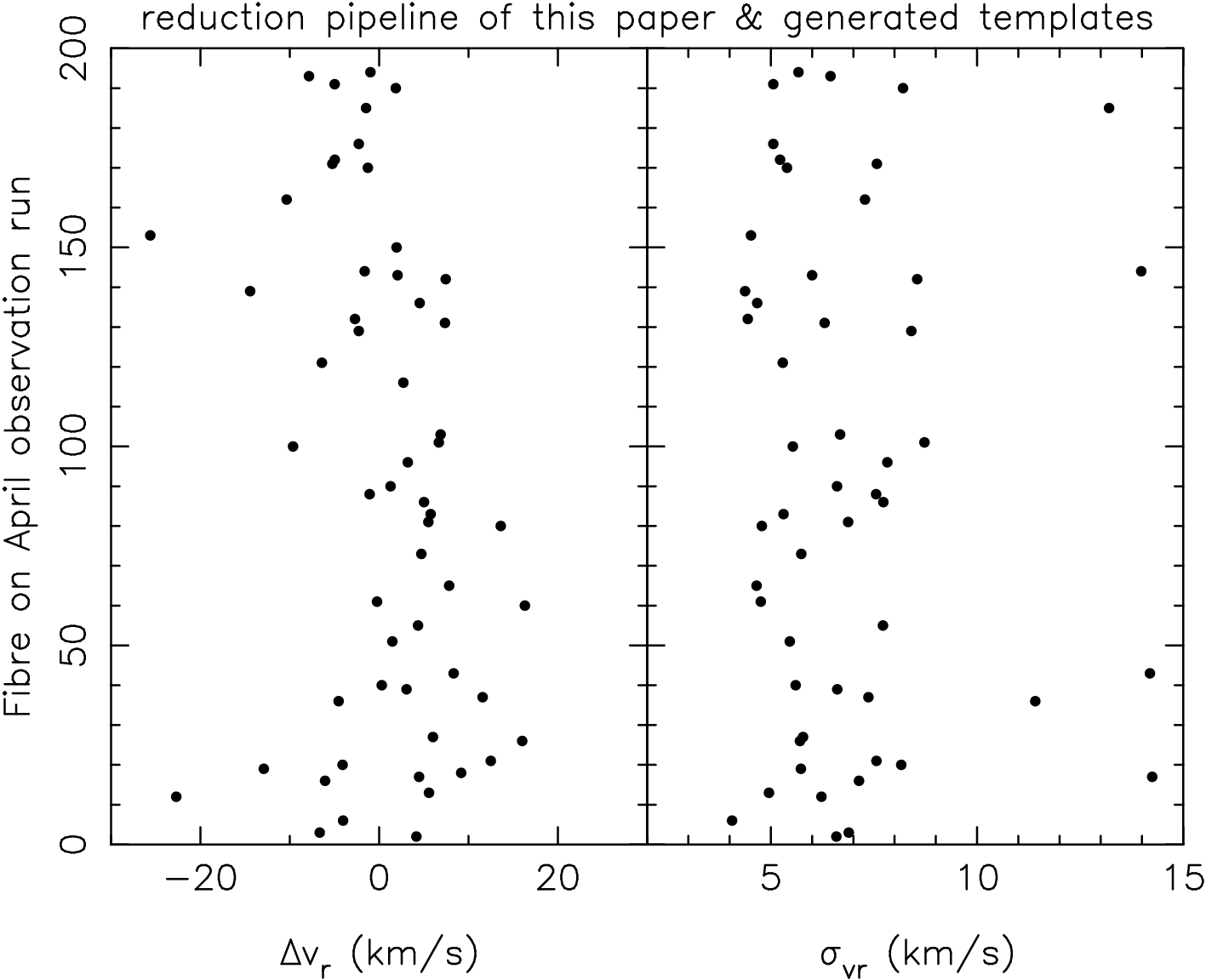}
\caption{Difference in radial velocities between the April and December observations when reducing the data using the
2dfdr package (left panels) and the reduction pipeline presented in this paper (right panels). For each set of panels,
the offset in radial velocity, $\Delta v_r=v_{r\mathrm{,\,Apr}}-v_{r\mathrm{,\,Dec}}$, has been plotted (left) along
with the uncertainties on these measurements (right). Using the reduction techniques developed in this paper produces
lower systematic errors with $\Delta v_r\sim0\,\kms$ and a lower internal dispersion of the offsets.}
\end{figure*}

To analyze the efficiency of the procedure we present here, we observed one of the fields of our Canis Major survey
twice. The first observation run was performed on 9th April 2004, using the spectrograph 1 settings (CCD1 and 1200V
grating; this is our April dataset) and the second run was performed on 3rd December 2004, this time using spectrograph
2 settings (CCD2 and 1200R grating; this is our December dataset). As we showed, the radial velocities derived from the
December observations on CCD2 should not have important systematic effects. Moreover, the use of the 1200R grating
produces low uncertainties on the velocities ($\sim 2\kms$). In the following, we will use the December radial
velocities as a reference and analyze the influence of our pipeline on the April radial velocities.


When using the 2dfdr package to reduce the April observations (left panels of Figure~9), the radial velocities are
broadly $\sim10-20\,\kms$ lower than the reference velocities measured in the December run, but some stars have offsets
as low as $-30\,\kms$ or as high as $20\,\kms$. At the same time, the uncertainties on these velocities are mainly in
the range $6-8\,\kms$ with, here as well, groups of outliers with uncertainties increasing to $\sim15\,\kms$ on
different portions of the CCD (for low fibres or fibres between 130 and 150).

In contrast, the reduction of the April observations on CCD1 using the pipeline described in this paper (right panels of
Figure~9) yields lower offsets compared to the reference velocities. The mean offset for all the CCD has changed from
$-9\pm10\kms$ to $0\pm8\kms$ and there are only a few outliers ($\sim2$). Similarly, the uncertainties on each radial
velocity are slightly lower and are more clustered, with no more groups of fibres producing outliers with high
uncertainties.

Though  the use  of the  reduction  pipeline described  in this  paper produces much more reliable  radial velocities,
low systematic offsets still  remain. On average,  this  effect  is  limited ($\sim5\,\kms$  and certainly  less  than
$10\,\kms$)  and  could  either  be due  to  the reduction pipeline  that may not completely correct  the LSF asymmetry
effects   (but  it   has   been   shown  that   there   are  no   more calibration-related issues) or  it could in fact
be  due to systematic errors in  the reference  velocities. We have shown that the  emission lines observed  through
CCD2 do not present any asymmetry  but what  remains unclear is the role that the variations of the width of the LSF of
CCD2 (see Figure~8) has on the radial velocities.

\section{Summary}

We have constructed an improved reduction pipeline for 2dF observations. The systematic
offsets that appeared between radial velocities measured from CCD1 data and those measured
from CCD2 data can now be corrected. It has
been shown that they are due to an asymmetric line spread function, which for the data
we analysed, particularly affects CCD1. This effect can be countered by:

\begin{enumerate}
\item modeling the asymmetric LSF by the sum of two Gaussian functions --- the first one represents the LSF of the system
if it were perfect and the second one is a perturbation of this system;
\item fitting the LSF model to emission lines of the calibration lamps for each group of 10 fibres;
\item using this LSF model to calibrate the corresponding spectra;
\item using this LSF model to downgrade high resolution template spectra so that the observations can be
cross-correlated with templates similarly deformed.
\end{enumerate}

The difference in  radial velocity in a set of Red Giant Branch stars that has been observed twice (once on each CCD)
is highly reduced when using this procedure  ($\sim\pm5\,\kms$). Moreover the radial velocity measurements show a more 
consistent behaviour throughout the CCD than when using only the AAO reduction pipeline.

Even though the 2dF spectrograph is to be decommissioned at the end of this year, archival data from the past decade
could be improved by the reduction procedure we propose here. In particular, studies of the kinematics of Galactic
populations could benefit highly from a reduction of the internal dispersion of the radial velocity datasets. The pipeline has only
been shown to work for a particular dataset, yet we suspect that observations obtained through other gratings or with
other central wevelengths and that present the same kind of systematic effects can be corrected in a similar way.

\section*{Acknowledgments}
We would like to thank Robert Sharp for useful comments on this paper.



\begin{thebibliography}{}
\bibitem[Bagnulo et al.(2003)]{bagnulo03}Bagnulo, S., Jehin, E., Ledoux, C., Cabanac, R., Melo, C., Gilmozzi, R., The
ESO Paranal Science Operations Team 2003, The Messenger, 114, 10
%
\bibitem[Cannon(2002)]{cannon02}Cannon R. 2002, AAO Newsletter No. 101, p22
%
\bibitem[Lewis et al.(2002)]{lewis02}Lewis I. J. et al. 2002, MNRAS, 333, 279
%
\bibitem[Martin et al.(2004)]{martin04}Martin, N.~F., Ibata R.~A., Conn, B.~C., Lewis, G.~F., Bellazzini, M., Irwin
M.~J., McConnachie, A. W. 2004, MNRAS, 355, L33
%
\bibitem[Martin et al.(2005)]{martin05}Martin, N.~F., Ibata R.~A., Conn, B.~C., Lewis, G.~F., Bellazzini, M., Irwin,
M.~J. 2005, MNRAS, submitted {\tt astro-ph/0503705}
%
\bibitem[Press et al.(1992)]{press92}Press, W.~H., Flannery, B.~P., Teukolsky, S.~A., Vetterling, W.~T. 1992, Numerical
Recipes. Cambridge Univ. Press, Cambridge
%
\bibitem[Stanford \& Cannon(2002)]{stanford02} 
      Stanford, L., Cannon, R. 2002, 2dF technical note, {\tt http://www.aao.gov.au/2df/technotes/fibvel.ps.gz}
%
\bibitem[Taylor et al.(1996)]{taylor96} 
       Taylor K., Bailey J., Wilkins T., Shortridge K., Glazebrook K. 1996, adass, 5, 195
%
\end{thebibliography}
\end{document}